\documentclass[aps,amsmath,floats,floatfix, twocolumn,
superscriptaddress,nofootinbib,reprint]{revtex4}

\usepackage{hyperref}

\usepackage{graphicx} 
\usepackage{xspace}
\usepackage[usenames,dvipsnames]{color}
\usepackage{amssymb}
\usepackage{longtable}

\newcommand{\Caltech}{\affiliation{Theoretical Astrophysics 350-17,
    California Institute of Technology, Pasadena, CA 91125, USA}}

\newcommand{\mb}[1]{{\mathbf#1}} 

\begin{document}

\title{
Geometrodynamics:  The Nonlinear Dynamics of Curved Spacetime
}

\author{Mark A.~Scheel} \Caltech 
\author{Kip S.~Thorne} \Caltech 

\date{May 30, 2014} 

\begin{abstract}
  We review discoveries in the nonlinear dynamics of curved spacetime,
  largely made possible by numerical solutions of Einstein's
  equations.  We discuss critical phenomena and
  self-similarity in gravitational collapse, the behavior of spacetime
  curvature
  near singularities, the instability of black strings in 5 spacetime
  dimensions, and the collision of four-dimensional black holes.  We
  also discuss the prospects for further discoveries in
  geometrodynamics via observation of gravitational waves.
\end{abstract}

\maketitle

\section{Introduction}

In the 1950s and 60s, John Archibald Wheeler \cite{geometrodynamics} speculated that curved, 
empty spacetime could exhibit rich, nonlinear dynamics --- which he called 
\emph{geometrodynamics} --- analogous to the writhing surface of the ocean 
in a storm.  Wheeler exhorted his students and colleagues 
to explore geometrodynamics by solving Einstein's general relativistic 
field equations.

In 1965, Yakov Borisovich Zel'dovich, with his young prot\'eg\'es 
Andrei Doroshkevich and Igor
Novikov \cite{DZN:1965}, gave strong evidence that, when a
highly deformed star collapses to form what would later be called a black
hole, the hole's curved spacetime, by its nonlinear dynamics, will 
somehow shake off all the deformations, thereby becoming a completely smooth, 
axially symmetric hole.  

The Wheeler/Zel'dovich challenge of exploring geometrodynamics in general, 
and for black holes in particular, has largely resisted analytic 
techniques.  In the 1980s and 90s, that resistance motivated the authors 
and our colleagues to 
formulate a two-pronged attack on geometrodynamics: \emph{numerical 
solutions of Einstein's equations} to discover general relativity's
predictions, and \emph{observations of gravitational waves} from
black-hole births and black-hole collisions 
to test those predictions.  The numerical
simulations have now reached fruition, and the observations will do so
in the near future.  

In this article---dedicated to the memory of Zel'dovich and Wheeler (who 
deeply respected each other despite cold-war barriers, and who were
primary mentors for one of us, Kip Thorne)---we shall 
present an overview of some of the most interesting things we have learned
about geometrodynamics from numerical simulations, and a preview of
what gravitational-wave observations may bring. More specifically:

We shall describe geometrodynamic discoveries in four venues: \emph{gravitational
implosion}, where a phase transition, discrete self-similarity and
critical behavior have been observed
(Sec.\ \ref{sec:Choptuik}); the dynamics of spacetime near \emph{singularities},
where richly chaotic behavior has been observed (Sec.\ \ref{sec:Singularities});
the unstable evolution of a \emph{black string} in five spacetime dimensions,
where a dynamical sequence of strings linking black holes has been observed
(Sec.\ \ref{sec:BlackString}); and \emph{collisions of black holes} in four
spacetime dimensions, where dynamical interactions of tidal 
tendexes and frame-drag vortexes have been observed (Sec.\ \ref{sec:BBH}).
Then we shall briefly describe the prospects for observing some of these
phenomena via \emph{gravitational waves} (Sec.\ \ref{sec:GW}).

\section{Gravitational Collapse: Phase Transition and Critical Behavior}
\label{sec:Choptuik}

The first numerical simulations to exhibit rich geometrodynamics were 
in 1993, by Matthew Choptuik \cite{Choptuik1993}, who was then a 
postdoc at the University of Texas.  Choptuik simulated the spherical
implosion (Fig.\ \ref{fig:WaveImplosion}) of a linear, massless scalar field 
(satisfying $\Box \Psi = 0$).  The field's 
energy, momentum and stress (which are quadratic in the field)
generated spacetime curvature, with which
the field interacted via the curvature's influence on its wave operator
$\Box$. That interaction produced surprising nonlinear dynamics:   

\begin{figure}[h!]
\begin{center}
\includegraphics[width=0.95\columnwidth]{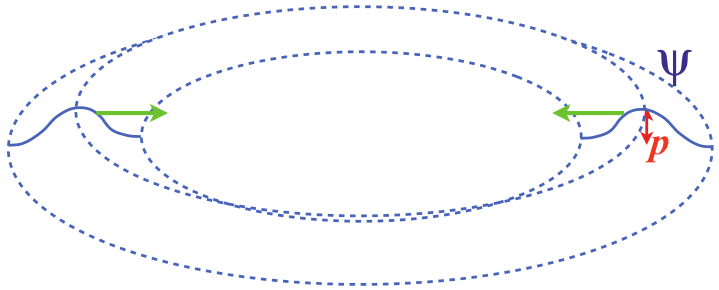}
\end{center}
\caption{The implosion of a scalar wave $\Psi$ with amplitude $p$ and
a particular waveform.}
\label{fig:WaveImplosion}
\end{figure}

If the wave amplitude $p$, for some chosen ingoing waveform, was larger than some critical
value $p_*$, the implosion produced a black hole.  If $p$
was smaller than $p_*$, the imploding waves passed through
themselves,  
travelled back outward, and dispersed. For $p=p_*$, the imploding waves
interacted with themselves nonlinearly (via their spacetime curvature),
producing a sequence of frequency doublings and wavelength halvings, 
with an intriguing, 
discretely self-similar pattern that was independent of the initial, 
ingoing waveform.  Waves with ever decreasing wavelength
emerged from the ``nonlinearly boiling'' field, 
carrying away its energy, and ultimately
leaving behind what appeared to be an infinitesimal naked singularity 
(a region with infinite spacetime curvature, not hidden by a 
black-hole horizon).
One year after Choptuik's simulations, the mathematician
Demetrios Christodoulou \cite{Christodoulou94} developed a 
rigorous proof that the endpoint, for $p=p_*$, was, indeed, a naked singularity.

The transition of the implosion's endpoint, from black hole for $p>p_*$,
to naked singularity for $p=p_*$, to wave dispersal for $p<p_*$, was
a phase transition analogous to those that occur in condensed-matter physics.
And, as in condensed matter, so also here, the phase transition exhibited
scaling:  For $p$  slightly larger than $p_*$, the mass of the final black hole scaled as $M_{\rm BH} \propto (p-p_*)^\beta$.  For $p$ slightly below $p_*$,
the radius of curvature 
of spacetime at the center of the ``boiling region'' reached a minimum value,
before field dispersal, that scaled as 
$\mathcal R_{\rm min}  \equiv 
(R^{\mu\nu\sigma\rho} R_{\mu\nu\sigma\rho})_{\rm max}^{-1/4} 
\propto (p_*-p)^\beta$, with the same numerically measured
exponent $\beta = 0.374$. Here $R_{\mu\nu\sigma\rho}$ is the Riemann
curvature tensor. 

Choptuik's discovery triggered many follow-on simulations.  Most interesting
to us was one by Andrew Abrahams and Charles Evans \cite{AbEv93}, later 
repeated with higher resolution by Evgeny Sorkin \cite{sorkin:2011}. 
In their simulations, the imploding, spherically symmetric scalar field was 
replaced by an imploding, axially symmetric (quadrupolar) gravitational
wave --- so they were dealing with pure, vacuum spacetime as envisioned
by Wheeler.  Once again, there was a critical wave amplitude $p_*$; and for 
$p$ near $p_*$ the behavior was similar to the scalar-wave 
case, to within numerical error:  for $p=p_*$, strong evidence for discretely 
self-similar evolution leading to an infinitesimal final singularity; 
for $p>p_*$, the same
black-hole mass scaling $M_{\rm BH} \propto (p-p_*)^\beta$; for 
$p<p_*$, the same spacetime-curvature scaling 
$\mathcal R \propto (p_*-p)^\beta$;
and to within numerical accuracy, the scaling exponent was the same:
$\beta = 0.38$ for the quadrupolar gravitational waves, and $\beta = 0.374$ for
the spherical scalar wave.  This is reminiscent, of course, 
of the universality one encounters in condensed-matter phase transitions.

For a detailed review of these and many other studies of critical
gravitational implosion, see an article by Carsten Gundlach \cite{gundlach}.

\section{Generic Space-Time Singularities}
\label{sec:Singularities}

\subsection{BKL Singularity}
\label{sec:BKL}

The singular endpoint of the implosions described above is \emph{non-generic},
in the sense that no singularity occurs if
$p_*$ is only infinitesimally different from $p$. 

However, there are other situations that lead to 
\emph{generic} singularities.\footnote{Perhaps the most important generic 
singularity for astrophysics is one that arises when matter with
negligible pressure is present.  
Leonid Petrovich Grishchuk in 1967 \cite{Grishchuk67} showed that the
matter, generically, undergoes gravitational collapse to form
flat pancakes with infinite density and curvature; and 
in 1970 Zel'dovich \cite{Zeldovich70} showed that, 
astrophysically, pressure
halts the collapse before the infinities but after the pancake structure 
has been strongly established.
A few years later Zel'dovich realized that these pancakes, seen edge on, 
are filamentary structures that 
astronomers observe in the distribution of galaxies on the sky.} 
This was demonstrated analytically in the 1960s by Roger Penrose,
Stephen Hawking and others, using tools from differential topology 
\cite{HawkingPenrose1969}.
In 1969--70, Vladimir Belinsky, Isaac Khalatnikov and Evgeny Lifshitz 
\cite{BKL70}
used approximate differential-geometry techniques to deduce the 
geometrodynamical 
behavior of spacetime as it nears one generic type of singularity, a type
now called \emph{BKL}.

In the 1970s, 80s and 90s, there was much skepticism in the US and UK 
about this BKL analysis, because its rigor was much lower than that of
the Penrose-Hawking singularity theorems.  (This lower rigor was inevitable,
because the geometrodynamical approach to a singularity is very complex---see
below---and deducing its details is much more difficult than proving that
a singularity occurs.)  As a result, the BKL geometrodynamics came to be
called, in the West, the \emph{BKL conjecture}.  

There was little hope 
for proving or disproving this ``conjecture'' by analytical techniques,
so the skeptics turned to numerical simulations for probing it.  After
nearly a decade of code development, David Garfinkle in 2003
\cite{Garfinkle2004}
carried out simulations
which demonstrated that Belinsky, Khalatnikov and Lifshitz had been correct.
The geometrodynamical evolution, approaching a BKL singularity, is 
as they predicted, though with one additional feature that they had missed:
a set of nonlocal \emph{spikes} in the spacetime curvature 
\cite{LimAnderssonGarfinklePretorius2009}.  

\begin{figure}[b!]
\begin{center}
\includegraphics[width=0.95\columnwidth]{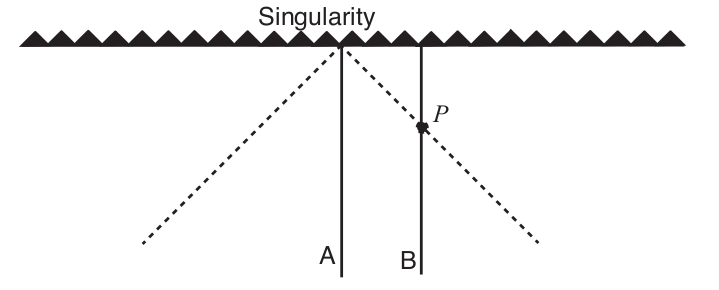}
\end{center}
\caption{The world lines of two observers, A and B, falling into a BKL
singularity (the solid lines), and the \emph{particle horizon} of observer A 
(dashed lines; a past light cone).  
Events outside the particle horizon can never influence observer A.}
\label{fig:BKL1}
\end{figure}

The BKL geometrodynamics can be described in terms of tidal-gravity 
observations by observers who fall into the BKL singularity on timelike
geodesics; Fig.\ \ref{fig:BKL1}. As two observers, A and B, approach the
singularity, they lose causal contact, in the sense that, after passing
through A's particle horizon (at point $P$ in the diagram), B can no longer
influence A. This causal decoupling is so strong, in the BKL spacetime, that
spatial derivatives cease having significant influence on the geometrodynamics,
as the singularity is approached---there is a \emph{spatial decoupling}---and
as a result, it turns out, there is no correlation between the late-time observations of
adjacent observers.

Each observer's experience, when approaching the singularity, can be 
described in terms of the tidal gravitational field $\mathcal E_{jk}$ that
he feels.  This tidal gravitational field has components, in the observer's
local Lorentz frame, that are equal to the space-time-space-time part of
the Riemann curvature tensor.
\begin{equation}
\mathcal E_{jk} = R_{j0k0}\;.
\label{eq:Ejk}
\end{equation}
The physical manifestation of the tidal field is a relative 
gravitational acceleration
\begin{equation}
\Delta a_j = - \mathcal E_{jk} \Delta x_k
\label{eq:Deltaa}
\end{equation}
of particles with vector separation $\Delta x_k$. (The tidal field acquires
its name from the fact that the Sun's and Moon's tidal field produces
the tides on the Earth's oceans.  In the Newtonian limit, the tidal field
is $\mathcal E_{jk}
= \partial^2 \Phi/\partial x_j \partial x_k$, where $\Phi$ is the Newtonian
gravitational potential.)  

Being a symmetric tensor, the tidal
field can be described by three orthogonal principal axes (unit vectors
$\mb e_{\hat 1}$, $\mb e_{\hat 2}$, $\mb e_{\hat 3}$), 
and their eigenvalues, $\mathcal E_{\hat j \hat j} \equiv \mb e_{\hat j}
\cdot \boldsymbol{\mathcal E} \cdot \mb e_{\hat j}$.  If $\mathcal E_{\hat 1
\hat 1} < 0$, then an object is tidally stretched along its 
$\mb e_{\hat 1}$ principal axis, and similarly for the other principal
axes.  If $\mathcal E_{\hat 1 \hat 1} > 0$, the object is tidally squeezed
along $\mb e_{\hat 1}$.  The tidal field, in vacuum, is trace-free, so
its eigenvalues must sum to zero, which means there must be a squeeze
along at least one principal axis and a stretch along at least one.

\begin{figure}[h!]
\begin{center}
\includegraphics[width=0.95\columnwidth]{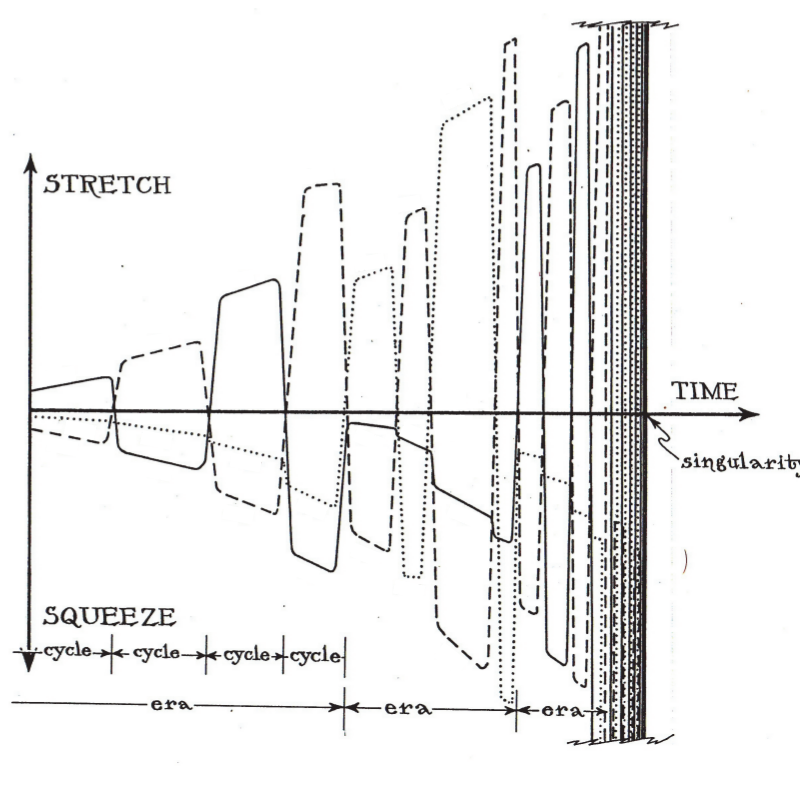}
\end{center}
\caption{The qualitative pattern of tidal stretches and 
squeezes experienced by an
observer when falling into a BKL singularity. The three eigenvalues
of the tidal field are plotted vertically as functions of time, with one
axis shown solid, another dashed, and the third dotted.  
Adapted from \cite{thorne:BlackHolesAndTimeWarps}} 
\label{fig:BKL2}
\end{figure}

Figure \ref{fig:BKL2} shows the pattern of stretches and squeezes experienced
by an observer when falling into the BKL singularity.  The pattern is divided,
in time, into \emph{cycles} and \emph{eras}.  During a single cycle, there
is a stretch along one axis and a squeeze along the other two.  Between
cycles, the stretch axis switches to squeeze and the more strongly squeezing
axis switches to stretch.  At the end of each era, the axes rotate in some
manner, relative to the observer's local Lorentz frame, and the pattern begins 
over.  The number of cycles in each era and the details of their dynamics
are governed by a continued-fraction
map that is chaotic, in the technical sense of chaos (extreme sensitivity
to initial conditions).  This chaos plays a key role in destroying correlations
between the observations of adjacent observers as they approach the 
singularity.  

The full details of this, as worked out by Belinsky, Khalatnikov and 
Lifshitz \cite{BKL70}, occasionally are violated (numerical simulations
reveal):
A cycle gets skipped, and during that skip, there is
an extreme spike in the tidal field that is more sensitive to spatial 
derivatives than expected, and whose details are not yet fully understood.
See \cite{LimAnderssonGarfinklePretorius2009} and references therein.   

\subsection{Singularities Inside a Black Hole}
\label{sec:BHSingularities}

This BKL behavior is speculated to occur in the core of a young black hole.
However, 
numerical simulations are needed to confirm or refute this speculation.

\begin{figure}[h!]
\begin{center}
\includegraphics[width=0.95\columnwidth]{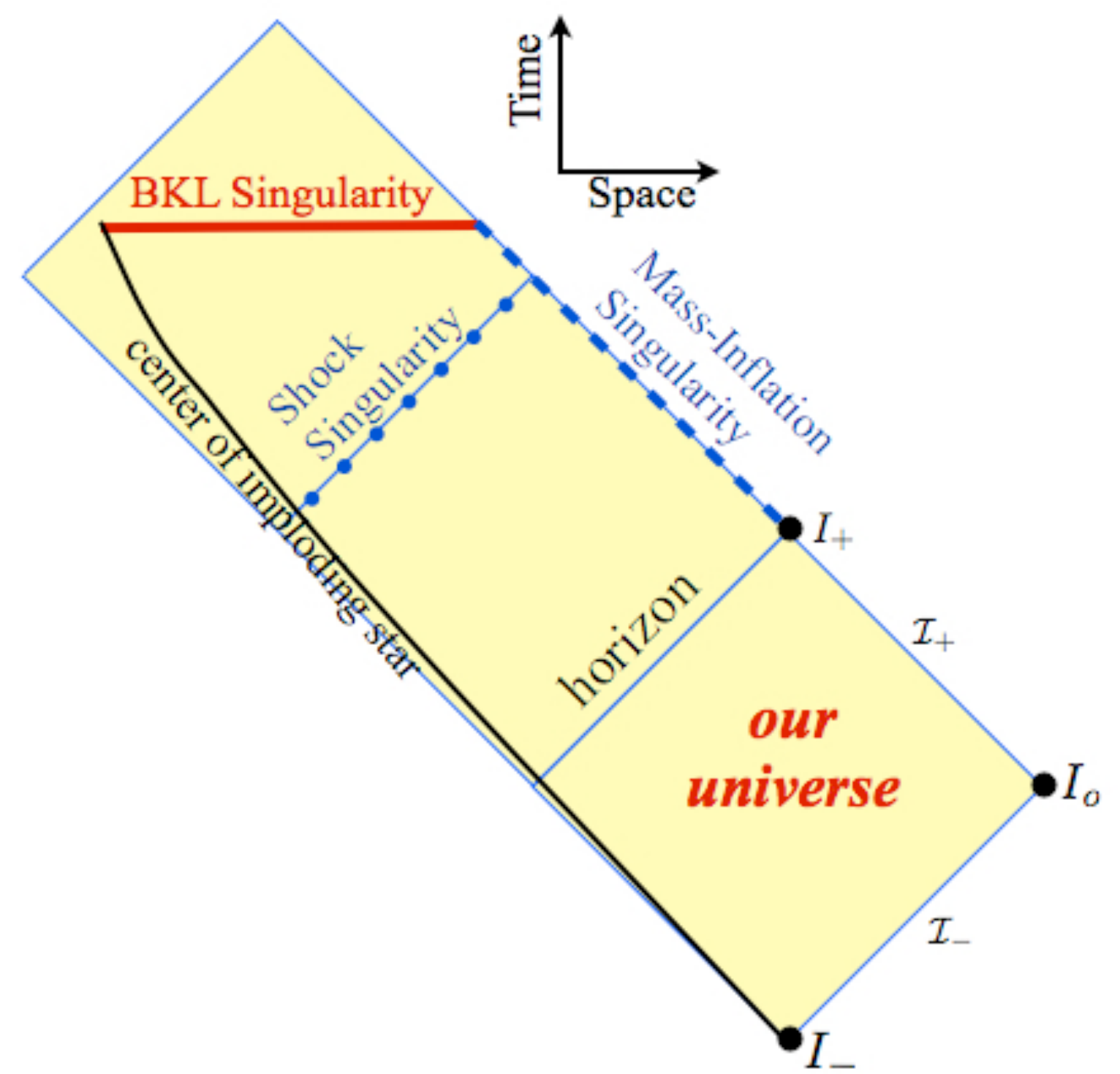}
\end{center}
\caption{Penrose diagram, depicting the causal structure outside and
inside an old black hole, as best we understand it today.  Ingoing and
outgoing null rays (hypothetical photons) travel along 45 degree lines,
and a conformal transformation has been used to compress our universe
into a finite diamond in the diagram.  The Kerr spacetime
is shaded.  The true spacetime is superposed on the Kerr spacetime; it
is the region bounded by the center of the imploding star (thin left
line), the BKL singularity (thick horizontal line), the mass-inflation
singularity (dashed line), and the infinities of our universe: future
timelike infinity labeled $I_+$, future null infinity labeled 
$\mathcal I_+$, spacelike infinity labeled $I_o$, past null infinity
labeled $\mathcal I_-$, and past timelike infinity labeled $I_-$.
It might well be that the true spacetime ends at the shock singularity, and
there is no BKL singularity beyond \cite{MarolfOri2012}.
} 
\label{fig:BKL3}
\end{figure}

As the black hole ages, the singularity in its core is speculated to 
break up into three singularities, one of BKL type, and two that are
much more  more gentle than BKL.  This speculation arises from the
expectation that, just as the hole's exterior spacetime geometry
settles down into the 
quiescent, axially symmetric state described by the Kerr metric, so its 
interior will settle down into the Kerr-metric state, except near two 
special 
regions called \emph{Cauchy horizons}, where the Kerr metric is 
dynamically unstable.  Nonlinear
geometrodynamics is thought to convert those Cauchy horizons into
\emph{null, generic singularities} (Fig.\ \ref{fig:BKL3}).

These singularities are null in the sense that ingoing or outgoing 
photons, in principle, could skim along them, not getting captured.  
The ingoing singularity (called a \emph{mass inflation singularity})
is generated, according to approximate 
analytical analyses \cite{Poisson90,Ori2000}, 
by radiation and material that fall into the black hole
and pile up along the ingoing Cauchy horizon, there gravitating intensely.
The outgoing singularity (called a \emph{shock singularity}
\cite{MarolfOri2012}) is
generated by ingoing radiation that 
backscatters off
the hole's spacetime curvature, and then travels outward, piling up
along the outgoing Cauchy horizon, there gravitating intensely.
In both cases, the piling-up stuff could be gravitational waves rather
than material or nongravitational radiation, in 
which case we are dealing with pure vacuum spacetime curvature---pure
geometrodynamics.

Neither of these null singularities is oscillatory, but at both of them,
the spacetime curvature goes to infinity (the radius of curvature
of spacetime $\mathcal R$ goes to zero). This divergence of curvature
happens so quickly at the shock singularity, that objects \emph{might} 
be able to pass through it, though with a destructive net compression
along two dimensions and net stretch along the third.  If so, 
they will likely then be destroyed in the BKL singularity.   

The mass-inflation singularity is expected also to
produce only a finite, not 
infinite, net compression and stretch of objects falling through.
If anything survives, its subsequent fate is totally unknown.

These (highly informed) speculations, which arise from extensive,
approximate analytically analyses, mostly perturbation theory, will
be tested, in the coming few years, by numerical simulations.  And
just as the BKL conjecture missed an important phenomenon (the
curvature spikes), so these speculations about the geometrodynamics
of black-hole interiors may fail in some modest, or even spectacular
way.  

For far greater detail on what we now know and speculate about the 
interiors of black holes and the bases for that knowledge and
speculation, see \cite{MarolfOri2012} and references therein.

\section{Black String in Five Spacetime Dimensions}
\label{sec:BlackString}

A remarkable example of geometrodynamics has been
discovered by Luis Lehner and Frans Pretorius \cite{LehnerPretorius:2010},
in numerical simulations
carried out in five spacetime dimensions.

Lehner and Pretorius began with a \emph{black string} in its
equilibrium state.  This black string is a vacuum solution of Einstein's
equations in five spacetime dimensions, with metric
\begin{eqnarray}
ds^2 &=& -\left(1-{2M\over r}\right) dt^2 + {dr^2\over 1-2M/r} 
\nonumber \\
&&+ r^2(d\theta^2 + \sin^2\theta d\phi^2) + dz^2\;.
\label{eq:BlackStringMetric}
\end{eqnarray}
This is precisely a four-spacetime-dimensional Schwarzschild black hole,
translated along the $z$ axis in the fifth (spatial) dimension.  The
event horizon is at $r=2M$; at fixed time $t$, it is a cylinder with
spherical cross section, $R \times S^2$.

In 1993, Ruth Gregory and Raymond LaFlamme 
\cite{GLF1993}
proved, analytically, that such a black
string is unstable against linear, axisymmetric 
perturbations with wavelength longer than about 
1.2 times the string's circumference. 
But little definitive was known about the instability's 
nonlinear, geometrodynamical evolution until Lehner and Pretorius's 
2010 simulations 
\cite{LehnerPretorius:2010}. They revealed that the string's horizon
evolves as depicted in Fig.\ \ref{fig:BlackString}:

\begin{figure*}
\begin{center}
\includegraphics[width=2.0\columnwidth]{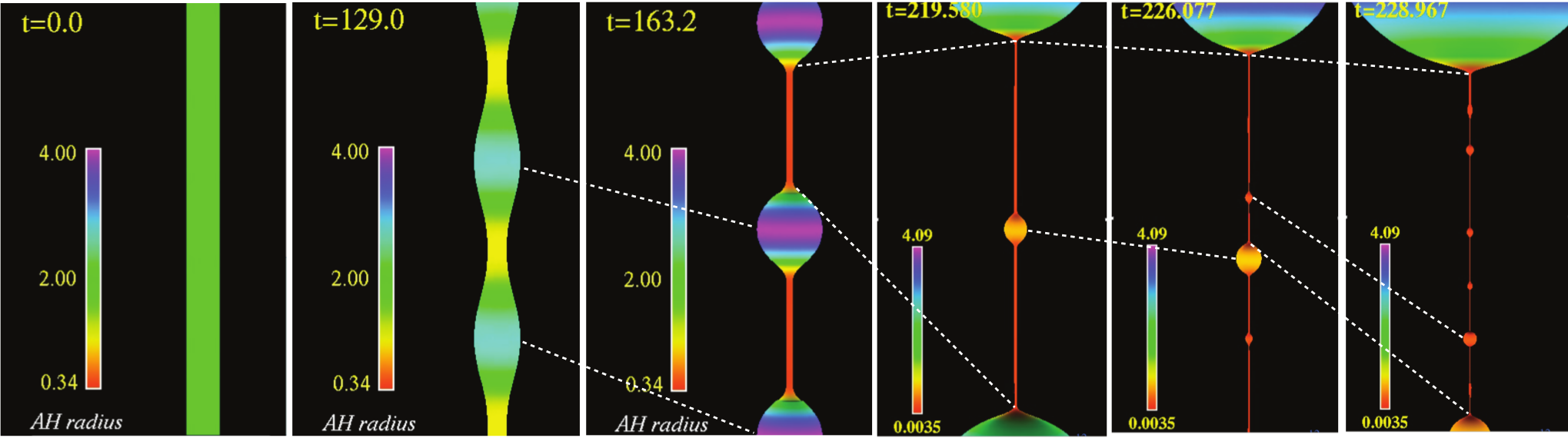}
\end{center}
\caption{(Color online.) A sequence of snapshots from a simulation of
the geometrodynamical evolution of a black string in 5 spacetime dimensions,
by Lehner and Pretorius \cite{LehnerPretorius:2010}. Each snapshot is an
embedding diagram of the black string's event horizon:  the horizon's 
intrinsic geometry is the same as the intrinsic geometry of the depicted
surface in flat space.  
}
\label{fig:BlackString}
\end{figure*}

The string develops a sausage instability---analogous to that of a 
magnetically confined plasma in a $Z$-pinch, 
and the Rayleigh-Plateau instability
for a cylinder of fluid confined by its surface tension---but 
with outgoing gravitational
waves carrying off energy. This instability gives rise to a chain of 
five-spacetime-dimensional black holes
linked by segments of shrunken black 
string---segments whose
circumferences are far smaller than that of the original string.  The
instability then repeats on each shrunken string, producing smaller black
holes linked by segments of more extremely shrunken string. Each successive
sausage instability produces its smaller black holes on an evolution timescale
proportional to the string's circumference.  These successively shorter
timescales converge:  An infinite sequence of instabilities, in
finite time, presumably ends in a naked singularity.

Because these simulations assumed 2-sphere symmetry from the outset,
we cannot be certain that their predictions are the black string's
true geometrodynamical evolution.  To learn the true evolution, we
need simulations in five spacetime dimensions, that do not assume any
symmetry.  Such simulations are beyond current capabilities, but may
be possible in a decade or so.  In the meantime, it is conjectured that
the Lehner-Pretorius evolution (Fig.\ 5) is the true evolution, because 
black strings have been proved stable against all {\it linear} 
nonspherical perturbations.

\section{Black-Hole Collisions}
\label{sec:BBH}


Recent advances in numerical simulations have
enabled the study of geometrodynamics in the violent collisions of two
black holes---including the generation of gravitational waves, and the
relaxation of the remnant to a single, quiescent, spinning black hole
(as predicted by Zel'dovich, Doroshkevich and Novikov \cite{DZN:1965}).  
We and collaborators
have developed a new set of so-called \emph{vortex/tendex tools} 
for visualizing this black-hole geometrodynamics~\cite{OwenEtAl:2011}.
We will first describe these tools, and then we will use them to
visualize the geometrodynamics of black-hole collisions.

\subsection{Vortex-tendex tools}
\label{sec:VortexTendex}

The gravitational field felt by a local observer is described by
the Riemann curvature tensor $R_{\mu\nu\sigma\rho}$.  
Any such observer, freely falling or accelerated, can decompose
Riemann into
an ``electric'' part,
$\mathcal E_{jk}$, defined by Eq.~(\ref{eq:Ejk}), and a ``magnetic''
part, $\mathcal B_{jk}$, defined by
\begin{equation}
\mathcal B_{jk} = \epsilon_{jpq}R_{pqk0}\;.
\label{eq:Bjk}
\end{equation}
Here the indices are components on the observer's local, orthonormal
basis, 
the index 0 refers to
the time component ({\it i.e.,} the component along the observer's
world line), Latin indices refer to the observer's three spatial 
components, and $\epsilon_{jpq}$ is the spatial Levi-Civita tensor.  Both
$\mathcal E_{jk}$ and $\mathcal B_{jk}$  are  symmetric and trace 
free in vacuum (the situation of interest to us).

As discussed in Section~\ref{sec:BKL},
$\mathcal E_{jk}$ is called the \emph{tidal field}, and
describes the tidal stretching and squeezing experienced by objects 
in the observer's local reference frame, according to
Eq.~(\ref{eq:Deltaa}).  The ``magnetic'' quantity $\mathcal B_{jk}$ is
called the \emph{frame-drag field}. The physical manifestation of this field
is a relative precession, or dragging of inertial frames: two gyroscopes
with vector separation $\Delta x_k$ will precess relative to each other
with angular velocity
\begin{equation}
\Delta \Omega_j = \mathcal B_{jk} \Delta x_k.
\label{eq:DeltaOmega}
\end{equation}
This  differential frame dragging is a general relativistic effect with no
analogue in Newtonian gravity.  Its global (non-differential) analog is
 one of the two relativistic effects
recently measured by Gravity Probe B~\cite{Everitt2011GPB}.  

Note that
the decomposition of the Riemann tensor into $\mathcal E_{jk}$ and
$\mathcal B_{jk}$ depends on the observer's reference frame.
Different observers at the same location, moving relative to each other,
will measure
different tidal fields and frame-drag fields.  This is the same
situation as for classical electromagnetism, in which the electromagnetic
tensor $F_{\mu\nu}$ can be decomposed into the familiar electric and
magnetic field vectors, $E_j=F_{j0}$ and $B_j=\epsilon_{jpq} F_{pq}$, 
in the same observer-dependent manner.  This
correspondence between gravitation and electromagnetism motivates the use
of the names ``electric'' and ``magnetic'' for $\mathcal E_{jk}$ and
$\mathcal B_{jk}$.

\begin{figure}[h!]
\begin{center}
\begin{picture}(0,110)(120,0)
\put(0,10){\includegraphics[width=0.49\columnwidth]{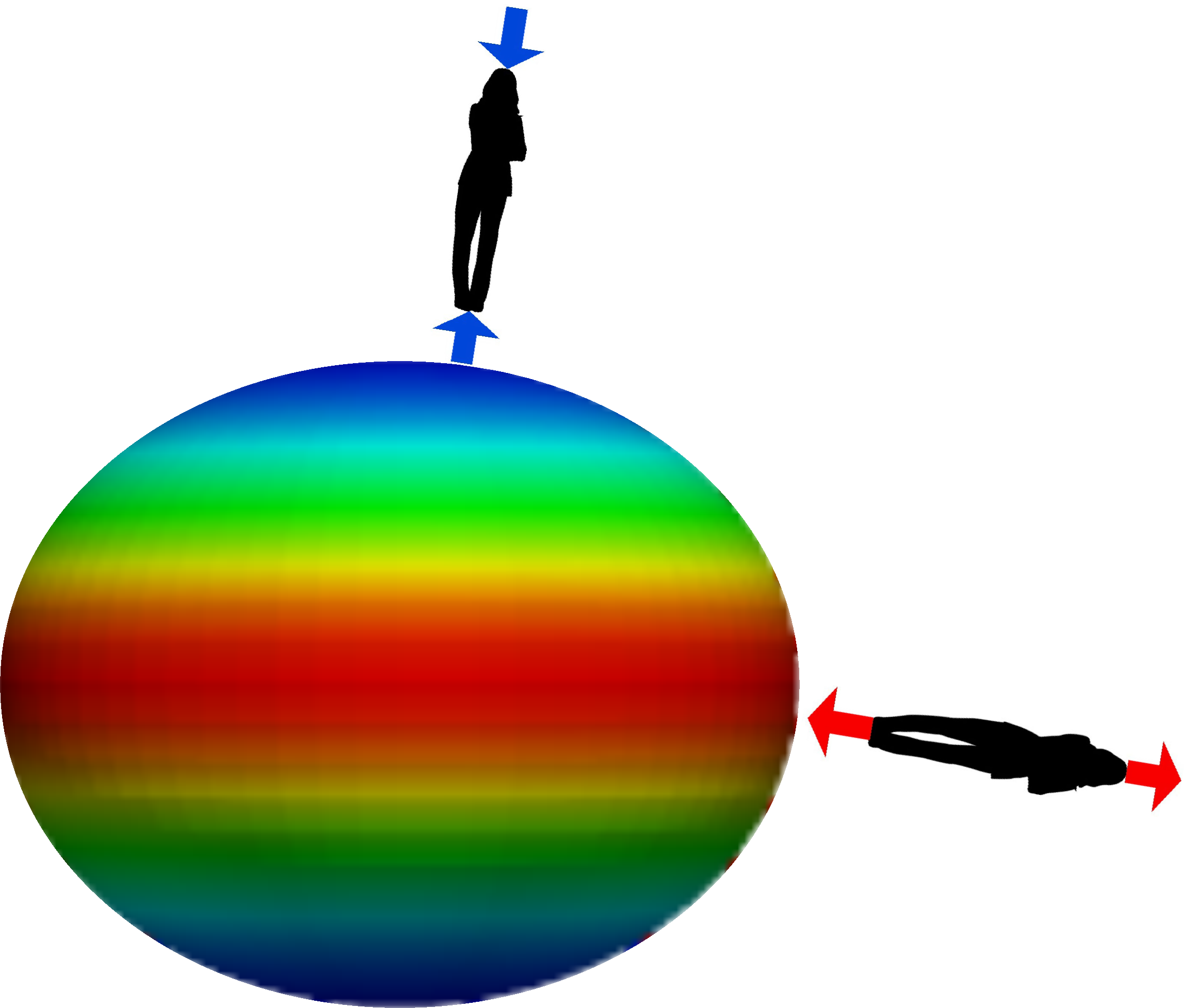}}
\put(155,-15){
\includegraphics[width=0.35\columnwidth]{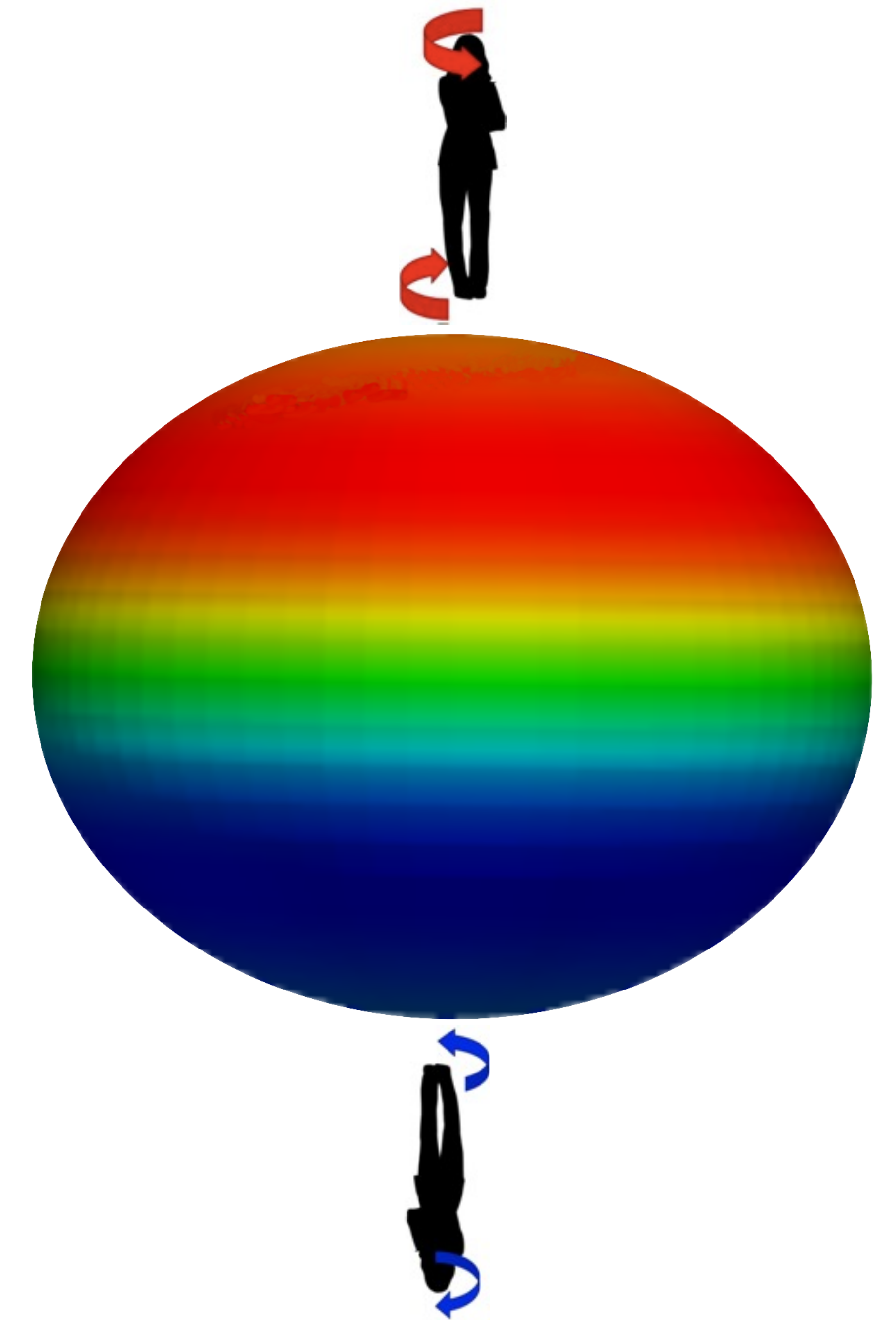}}
\end{picture}
\end{center}
\caption{(Color online.) A spinning black hole. {\bf Left}: Colors represent horizon tendicity 
$\mathcal E_{nn}$. There is a positive (blue or light gray) tendex on 
each of the
poles, and a negative (red or dark gray) tendex on the equator.  
Other (green or medium gray) regions have
small tendicity.
{\bf Right:} Colors represent horizon vorticity $\mathcal B_{nn}$.
There is a negative (red or dark gray) vortex on the north pole, 
and a positive (blue or light gray)
vortex on the south pole.  Other (green or medium gray)
regions have small vorticity.
The spin vector points out of the north pole.
\label{fig:HorizonTendicityVorticity}
}
\end{figure}

The frame-drag field $\mathcal B_{jk}$ and the tidal field 
$\mathcal E_{jk}$ are useful in describing
geometrodynamics at the surface of a black hole (its event horizon).  
If $\mb n$ is a unit
vector normal to the hole's horizon, with spatial components
$n^i$, then we define the {\em horizon tendicity} $\mathcal E_{nn}
\equiv n^j n^k \mathcal E_{jk}$ as the normal-normal component of the
tidal field.  For positive horizon tendicity, an object is tidally squeezed
normal to the horizon, and for negative horizon tendicity, an object is
tidally stretched normal to the horizon.  This is illustrated in the left
panel of
Fig.~\ref{fig:HorizonTendicityVorticity}.  We call a region on the horizon
with large tendicity a {\em horizon tendex}.  

We can similarly define the {\em horizon vorticity} $\mathcal B_{nn}
\equiv n^j n^k \mathcal B_{jk}$ as the normal-normal component of the
frame-drag field at the surface of a black hole. For positive horizon
vorticity, an object experiences a clockwise twist; for negative
horizon vorticity, the twist is counterclockwise.  We call a region on the
horizon with large vorticity a {\em horizon vortex}.  The horizon vorticity of
a spinning black hole is illustrated in the right panel of
Fig.~\ref{fig:HorizonTendicityVorticity}.

We now turn to vortex/tendex tools 
in regions away from the horizon.
In Section~\ref{sec:BKL} we discussed how, at any point, $\mathcal E_{jk}$
can be described by three orthogonal eigenvectors 
(unit vectors $\mb e_{\hat 1}$, $\mb e_{\hat 2}$, $\mb e_{\hat
  3}$), and their eigenvalues, $\mathcal E_{\hat j \hat j} \equiv \mb
e_{\hat j} \cdot \boldsymbol{\mathcal E} \cdot \mb e_{\hat j}$.
We call the eigenvalue $\mathcal E_{\hat j \hat j}$ the {\em tendicity}
associated with the corresponding eigenvector $\mb e_{\hat j}$; the
tendicity measures the tidal stretching or squeezing of an object
oriented along the eigenvector. 
In analogy with electric field lines, we define {\em tendex lines} as
the integral curves along each of the three eigenvectors $\mb e_{\hat
  j}$.  Whereas in electromagnetism there is a single electric field
line passing through each point, in geometrodynamics there are
generically three tendex lines passing through each point: one tendex
line for each of the three eigenvectors of $\mathcal E_{jk}$.  Because
(in vacuum) $\mathcal E_{jk}$ has vanishing trace, the eigenvalues must sum to
zero, so generically there exist both positive and negative tendex
lines at each point.  

\begin{figure}[h!]
\begin{center}
\begin{picture}(0,110)(120,0)
\put(0,0){\includegraphics[width=0.47\columnwidth]{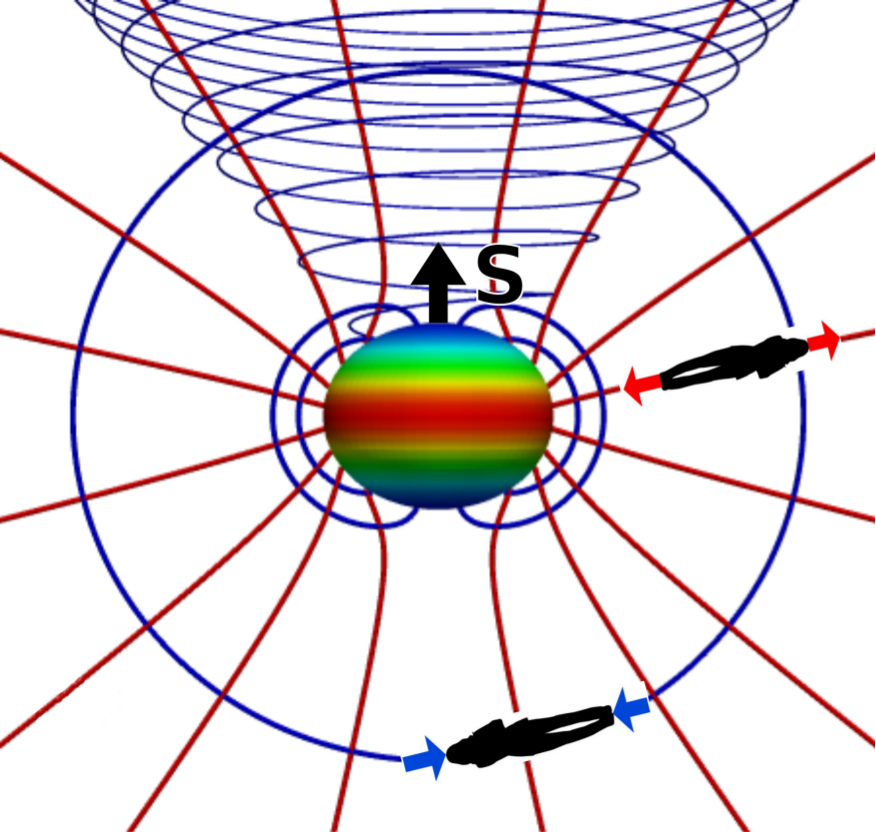}}
\put(120,0){
\includegraphics[width=0.47\columnwidth]{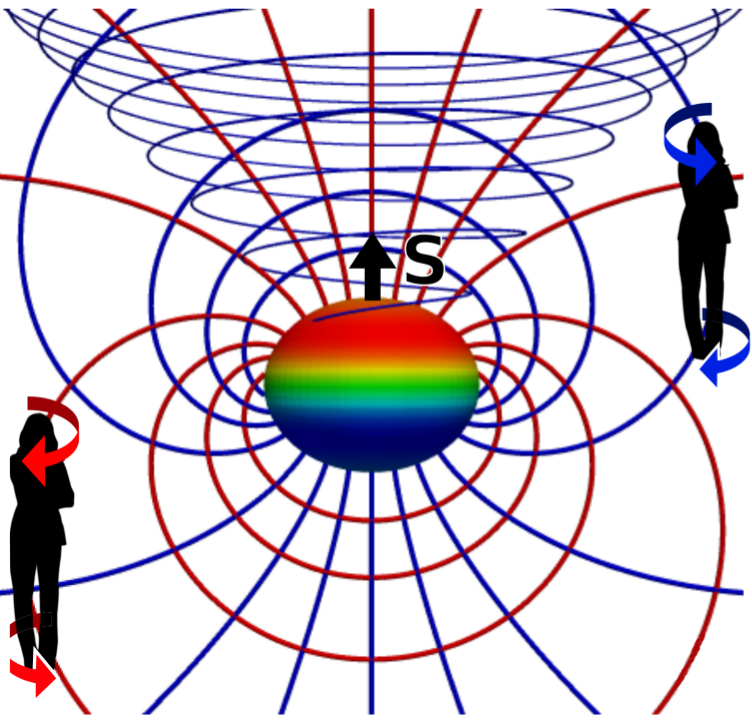}}
\end{picture}
\end{center}
\caption{(Color online.) 
Tendex lines (left panel) and vortex lines (right panel) near a 
spinning black hole.  Lines with positive eigenvalues (tidal squeeze or
clockwise twist)
are shown blue or light gray, and
lines with negative eigenvalues (tidal squeeze or counterclockwise twist) 
are shown red or dark gray.
At each point in space there are three intersecting tendex lines and three
intersecting vortex lines.
\label{fig:SingleBHTendexVortexLines}
}
\end{figure}

The left panel of Fig.\  
\ref{fig:SingleBHTendexVortexLines} shows the tendex lines outside
a rapidly rotating black hole.  A collection of tendex lines with particularly
large tendicity is called a {\em tendex}.  The rotating hole has a fan-shaped, 
stretching (red or dark gray) 
tendex sticking out of its equatorial horizon tendex, and a poloidal
squeezing (blue or light gray) 
tendex that emerges from its north polar horizon tendex, and reaches
around the hole to its south polar horizon tendex.

As with $\mathcal E_{jk}$, the  frame-drag field $\mathcal B_{jk}$ can be
described by three orthogonal eigenvectors and their eigenvalues.
An integral curve of one of these eigenvectors is called a {\em vortex
line}, and the corresponding eigenvalue is the {\em vorticity} associated
with that vortex line.  The vorticity of
a vortex line describes the twist, or differential frame dragging,
experienced by an object oriented along that vortex line: positive
vorticity corresponds to a clockwise twist, and negative vorticity
corresponds to a counterclockwise twist.

The vortex lines associated with a spinning black hole are shown
in the right panel of Fig.~\ref{fig:SingleBHTendexVortexLines}.  A counterclockwise
(red or dark gray) 
{\em vortex} (collection of large-negative-vorticity lines) emerges 
from the horizon's north polar vortex, reaches around the south polar 
region and descends back into the horizon's north polar vortex.
Similarly, a clockwise (blue or light gray) vortex emerges from the south polar 
horizon vortex,
reaches around the north polar region and descends back into the south 
horizon vortex.

\begin{figure}[h!]
\begin{center}
\begin{picture}(0,120)(120,0)
\put(0,0){
\includegraphics[width=0.95\columnwidth]{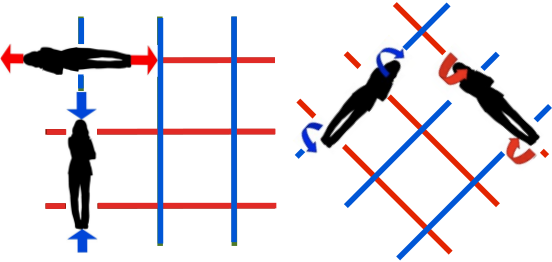}
}
\end{picture}
\end{center}
\caption{(Color online.)
  Snapshot of vortex and tendex lines for a gravitational plane wave
  traveling into the page.  Two orthogonal sets of tendex lines
  (left) are oriented 45 degrees with respect to two orthogonal sets
  of vortex lines (right).  The third set of vortex (and tendex) lines
  is normal to the page, with zero vorticity (tendicity).
\label{fig:PlaneWave}
}
\end{figure}

The black holes of Figs.\ \ref{fig:HorizonTendicityVorticity} 
and \ref{fig:SingleBHTendexVortexLines} 
have stationary (time-independent) vortex and tendex structures. 
Vortex and tendex lines, and their associated vortexes and tendexes, can
also behave dynamically. The equations of motion for $\mathcal E_{jk}$
and $\mathcal B_{jk}$ are similar to Maxwell's equations. Like
Maxwell, they have wavelike solutions---gravitational waves---in which
energy is fed back and forth between $\mathcal E_{jk}$ and $\mathcal B_{jk}$.
Figure~\ref{fig:PlaneWave} shows 
vortex and tendex lines for a plane gravitational wave without any
nearby sources.  As the wave passes an observer, the tendicities
and vorticities oscillate in sign with one oscillation per
gravitational wave period, leading to an oscillatory
stretch and squeeze in horizontal and vertical directions,
and an oscillatory twist in diagonal directions, out of phase with the
stretch and squeeze.

\subsection{Black-hole collisions}
\label{sec:BHcollisions}

We shall illustrate the geometrodynamical richness of black-hole collisions
by describing the vortex/tendex behaviors in
several numerical simulations. All these simulations were performed by 
members of the 
Collaboration to Simulate Extreme Spacetimes (SXS), which included
numerical relativists from Caltech, Cornell, the Canadian Institute for
Theoretical Astrophysics, and Washington State University at the time
of these simulations, and has since been expanded.  We are members of
this collaboration, and one of us (Scheel)
played a significant role in most of these simulations.

\subsubsection{Head-on collision of two black holes with transverse spins}

\begin{figure}[b!]
\begin{center}
\begin{picture}(0,170)(120,0)
\put(0,110){
\includegraphics[width=0.95\columnwidth]{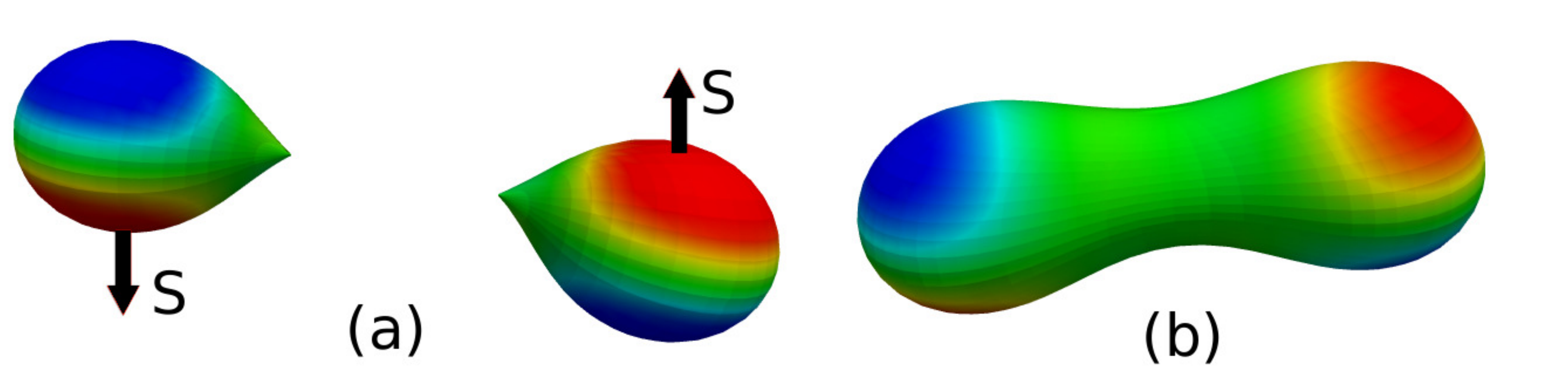}
}
\put(0,0){
\includegraphics[width=0.48\columnwidth]{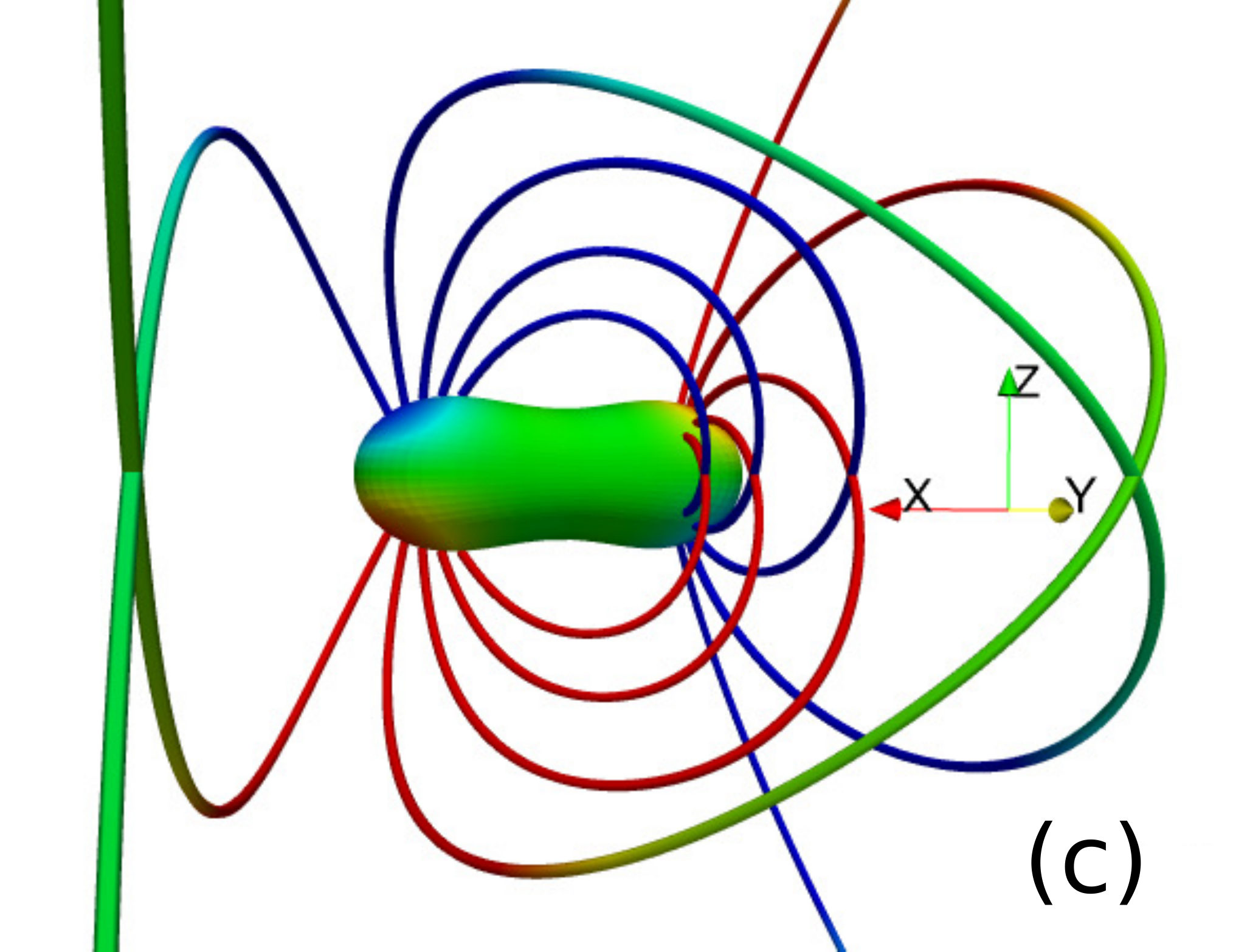}
}
\put(130,-10){
\includegraphics[width=0.45\columnwidth]{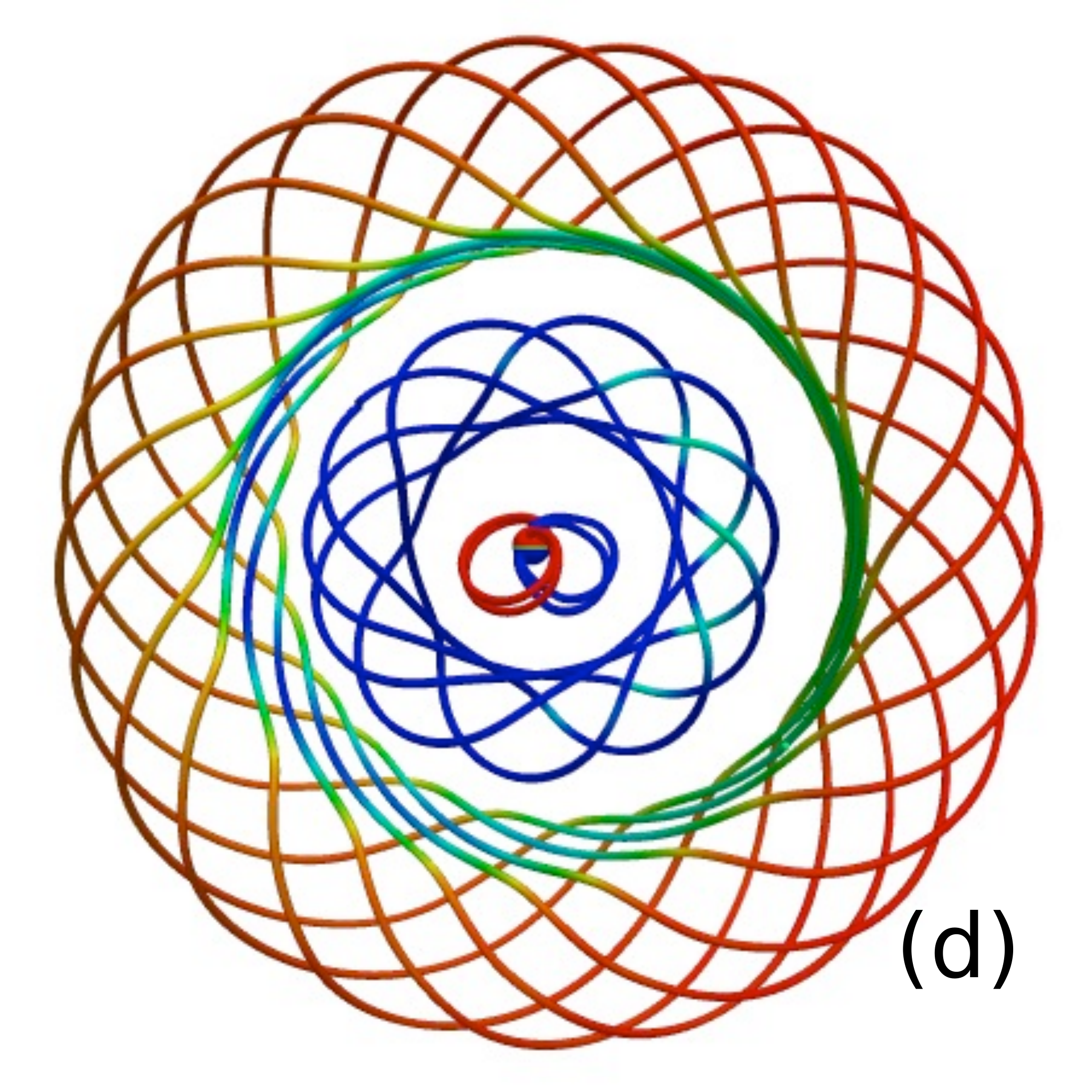}
}
\end{picture}
\end{center}
\caption{(Color online). The event horizons and vortex lines of 
spinning black holes, colliding head-on with transverse spins; 
from the simulation reported 
in~\cite{Lovelace:2009}. (a) Horizon vortexes and spin directions just 
before merger. (b)
Horizon vortexes just after merger, which retain their individuality.
(c) Vortex lines linking horizon vortexes of same
polarity (red to red or dark gray to dark gray; blue to blue
or light gray to light gray). Lines are color coded by vorticity
(different scale from horizon). (d) Sloshing of near-zone vortexes
generates toroidal vortex loops (two shown here) composed of orthogonal
vortex lines, traveling outward as gravitational waves; these
are accompanied by intertwined tendex lines (not shown). 
The horizon, with attached vortex lines, is
visible in the center. 
Figure adapted from~\cite{OwenEtAl:2011}.
See~\cite{movie:headon} for a movie of horizon vortexes for
this simulation.
\label{fig:HeadonVortex1}
}
\end{figure}

Our first example is a simulation~\cite{Lovelace:2009} of 
the head-on collision of two transversely spinning black holes,
depicted in Fig.~\ref{fig:HeadonVortex1}. 

As the black holes merge, the vortexes retain their individuality.  
When the four vortexes (one pair from each hole) discover each other, 
all attached to the same
horizon, they begin to interact: each one tries to convert those adjacent
to it into a replica of itself.  As a result, they exchange vorticities;
each oscillates back and forth between clockwise and counterclockwise.  
At the moment when all are switching direction, so the horizon 
momentarily has zero vorticity, the vortex lines have popped off the 
horizon and joined onto
each other, creating a toroidal structure, much like a smoke ring, that
is beginning to travel outward.  Simultaneously, adjacent to the
horizon much of the oscillation energy is stored in tendexes, which then
regenerate the horizon vortexes, but with twist directions reversed.
As the toroidal bundle of vortex lines travels outward, its motion 
generates toroidal tendex lines, intertwined with the vortex lines in
just such a manner as to become, locally, the gravitational-wave structure
described in Fig.\  \ref{fig:PlaneWave} above.

The process repeats over and over, with successive, toroidal, tendex/vortex
structures being ejected and morphing into gravitational waves.  The
waves carry away oscillation energy, and some oscillation energy goes
down the hole, so the oscillations die out, with an exponentiation time
of order an oscillation period.

\subsubsection{Collision of identical, spinning black holes, in an inspiraling 
circular orbit}

For collisions of orbiting (i.e. non head-on) black holes, the vortex
and tendex lines similarly travel to the wave zone and become
gravitational waves.  

\begin{figure}[b!]
\begin{center}
\begin{picture}(0,80)(120,0)
\put(0,0){\includegraphics[width=0.48\columnwidth]{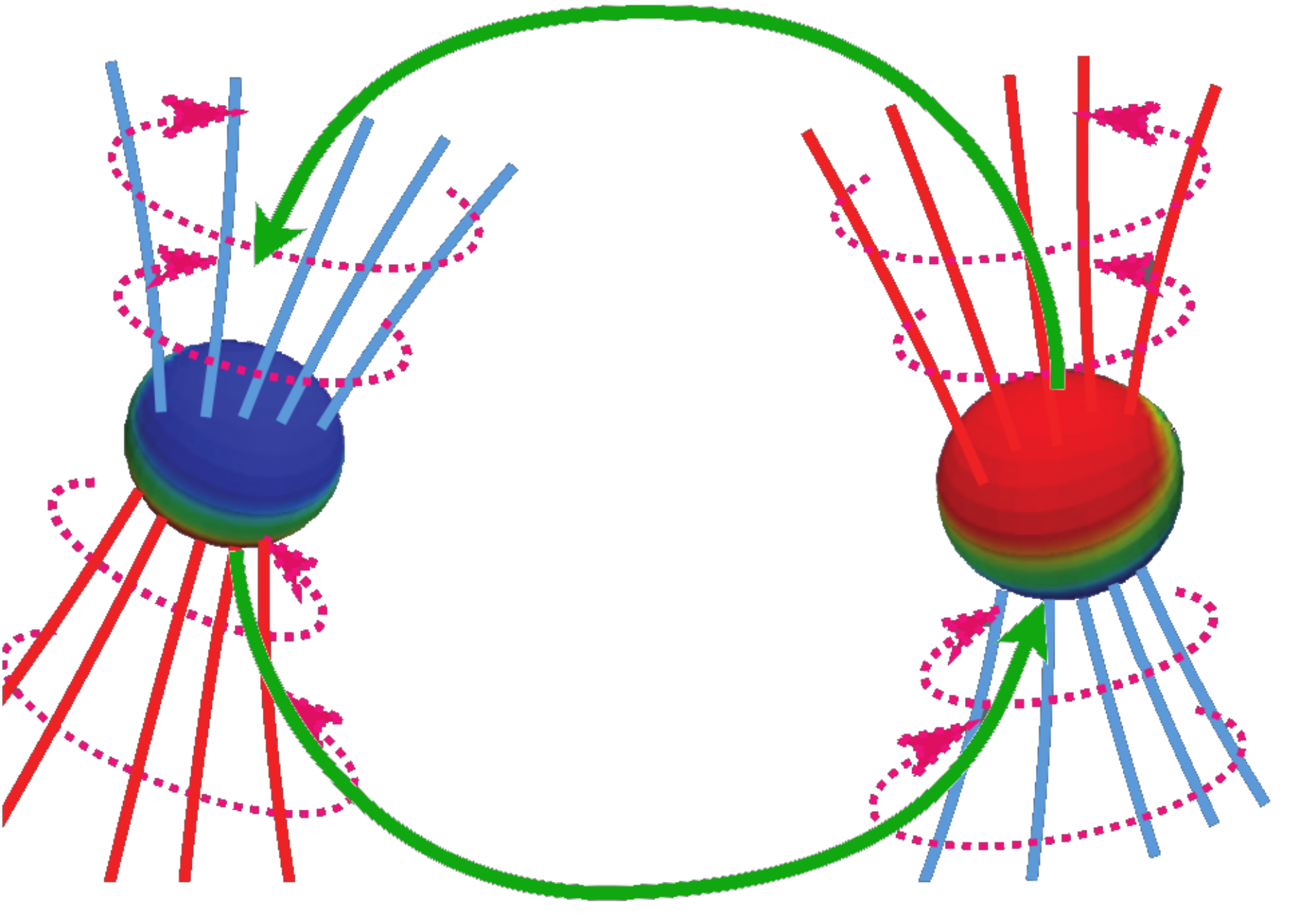}}
\put(150,0){\includegraphics[width=0.35\columnwidth]{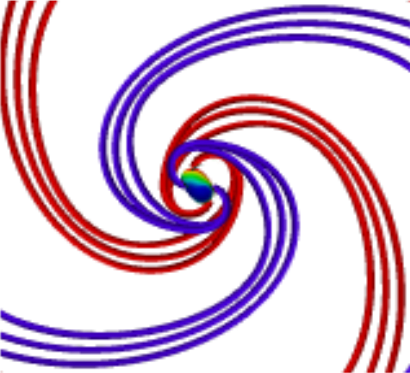}}
\end{picture}
\end{center}
\caption{(Color online.) {\bf Left:} 
Schematic of vortex lines and horizon vortexes
for two orbiting, spinning black holes about to merge. 
{\bf Right:} Schematic of 
vortex lines of the remnant black hole just after
merger, showing vortex lines extending to large distances; 
the entire pattern is rotating counterclockwise.
\label{fig:Sprinklerhead}
}
\end{figure}

The left panel of
Fig.~\ref{fig:Sprinklerhead} shows a schematic diagram of the horizon
vortexes and the vortex lines for the collision of two orbiting,
spinning black holes that are about to merge.  Just after merger, the
horizon vortexes retain their individuality, and travel around the horizon
of the merged black hole, trailing their vortex lines
outward and backward in a pattern like water from a spinning sprinkler head 
(shown schematically
on the right of Fig.~\ref{fig:Sprinklerhead}).
In the wave zone, the vortex lines acquire tendex lines and become
gravitational waves.  

Similarly, the near-field
tendex lines, attached to the merged hole's horizon tendexes, 
trail backward and outward in
a spiral pattern, acquiring accompanying vortex lines as they travel, and
becoming gravitational waves.  These horizon-tendex-generated 
gravitational waves have the opposite parity from the horizon-vortex-generated
waves; and there is a remarkable duality between the two sets of waves
\cite{Nichols:2012jn}

\begin{figure}[t!]
\begin{center}
\includegraphics[width=0.49\columnwidth]{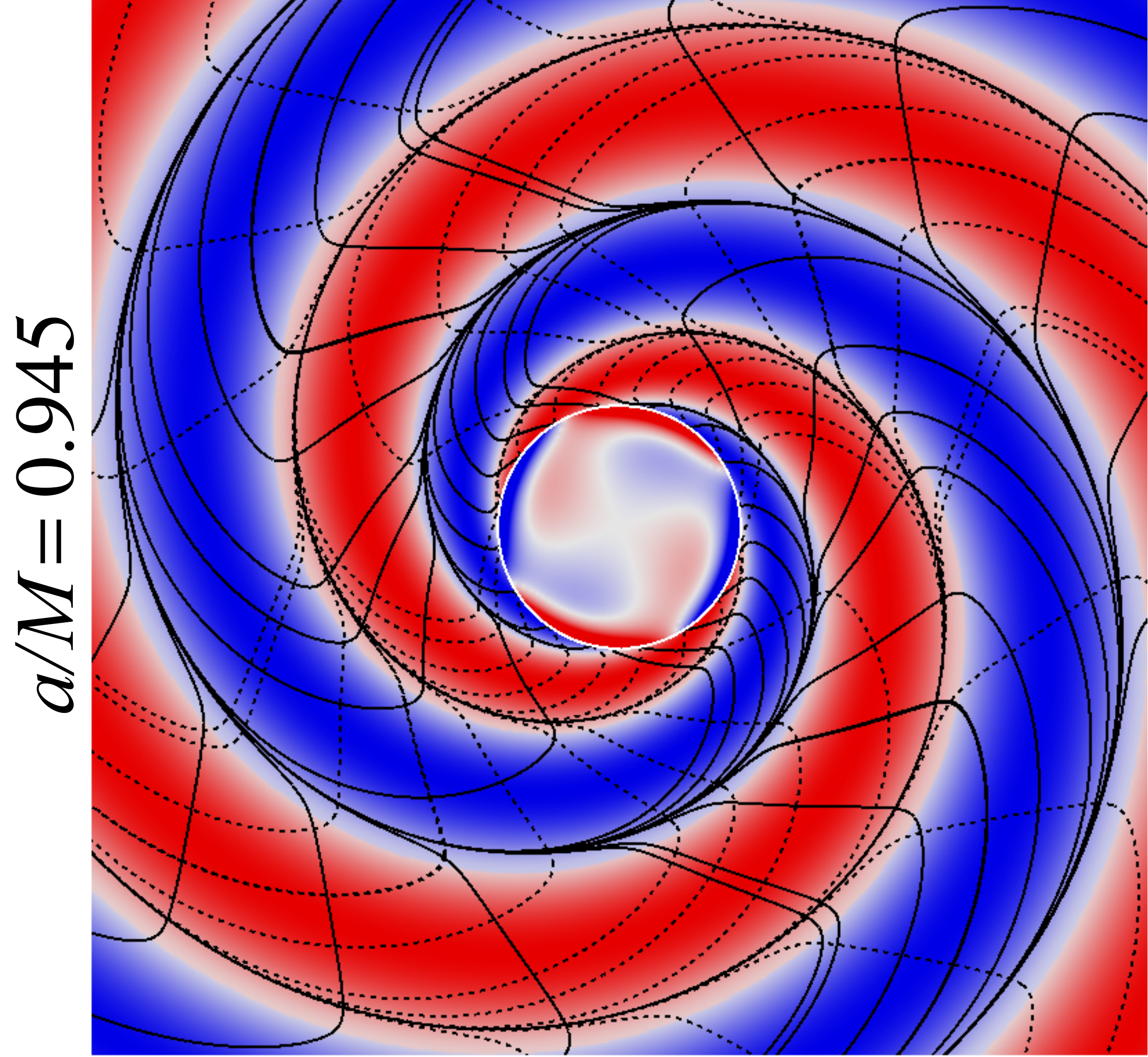}
\includegraphics[width=0.45\columnwidth]{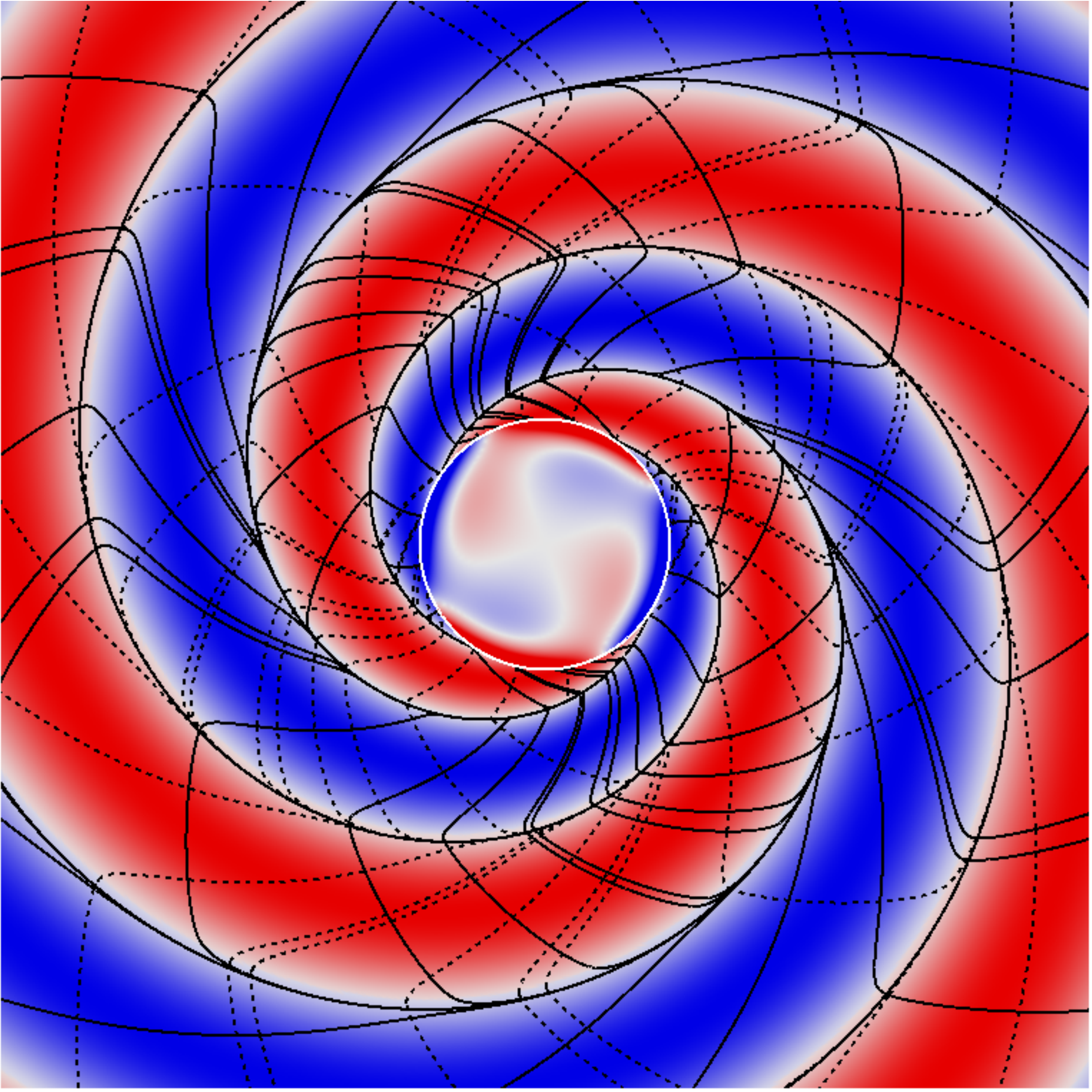}
\end{center}
\caption{(Color online.) 
Vortex and tendex structures for deviations from the final,
Kerr black hole, at late times after the merger of a black-hole binary.
These are the structures in the final hole's equatorial plane, and the
final hole has a dimensionless spin $S/M^2 = a/M =0.945$.
{\bf Left:} Perturbations generated by horizon vortexes---vortex lines 
(solid black for clockwise, dashed for 
counterclockwise), and vorticity of the dominant vortex at each point
(colored blue or light gray for clockwise and red or dark gray
for counterclockwise).
{\bf Right:} Perturbations generated by horizon tendexes---tendex lines
(solid black for squeeze, dashed for stretch), and tendicity of the
dominant tendex (blue or light gray for squeeze and red or
dark gray for stretch). Adapted from
Fig.\ 12 of \cite{Nichols:2012jn}.  
\label{fig:QNMKerr}
}
\end{figure}

Figure \ref{fig:Sprinklerhead} is schematic.  For a more precise depiction,
we focus on the merged hole at late times, when it is weakly perturbed
from its final, Kerr-metric state, and its perturbations are predominantly
$\ell=2,\; m=2$ quasinormal modes \cite{Nichols:2012jn}.  
Then the perturbations of the frame-drag
field, $\delta \mathcal B_{ij}$, generated by the horizon vortexes,
have the vortex lines and vorticities shown in the left panel of 
Fig.\ \ref{fig:QNMKerr}; and the perturbations of the tendex field, 
$\delta \mathcal E_{ij}$, generated by the horizon tendexes, have the
tendex lines and tendicities show in the right panel.  Notice that 
the two figures are very nearly identical, aside from sign (interchange
of red and blue, or dark and light gray).  
This is due to the (near) duality between the two sets
of perturbations.

\subsubsection{Extreme-kick black-hole collision}

An interesting example of geometrodynamics, and of the interplay
between vortexes and tendexes, is the ``extreme-kick''
black-hole collision first simulated, not by our SXS collaboration, but by 
Campanelli et\ al.\ \cite{Campanelli2007} and
others~\cite{Gonzalez2007b,LoustoZlochower:2010}. Our collaboration
has repeated their simulations, in order to extract the vortex and tendex
structures.

In these simulations,
two identical black
holes merge from an initially circular orbit, with oppositely directed
spins lying in the orbital ($x,y$) plane.  The name ``extreme-kick''
arises because gravitational waves
generated during the merger carry linear momentum preferentially in
either the $+z$ or $-z$ direction, resulting in a gravitational recoil
of the remnant black hole with speeds as high as thousands of km/s.  The
magnitude and direction of the recoil depends on the angle between 
holes' identical spin axes, and the holes separations, at the moment
of merger.  This angle
can be fine-tuned (for instance, to produce maximum recoil in the $+z$
direction) by adjusting the initial conditions in the simulation.

\begin{figure}[b!]
\begin{center}
\begin{picture}(0,110)(120,0)
\put(0,0){\includegraphics[width=0.47\columnwidth]{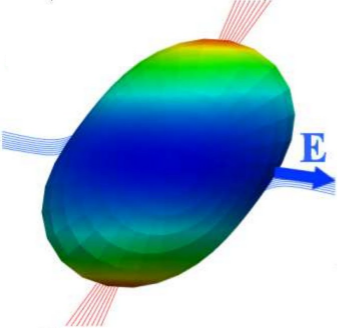}}
\put(120,0){
\includegraphics[width=0.47\columnwidth]{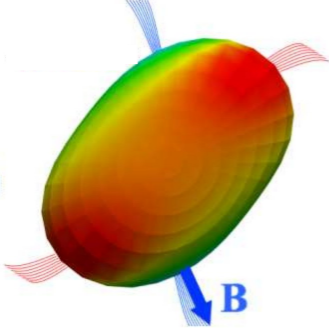}}
\end{picture}
\end{center}
\caption{(Color online.)
Remnant horizon, shown in the $xy$ plane,
just after merger, for a superkick simulation tuned
for maximum remnant recoil in the $+z$ direction; from
a simulation presented in~\cite{OwenEtAl:2011}.  
The black hole and vortex/tendex structures rotate counterclockwise. 
{\bf Left:} Colors show horizon tendicity; tendex lines are shown emerging
from the strongest horizon tendexes. {\bf Right:} Colors show horizon vorticity; vortex
lines are shown emerging from the strongest horizon vortexes. 
Figure adapted from~\cite{OwenEtAl:2011}. See~\cite{movie:superkickv} 
and~\cite{movie:superkickt} for movies of horizon vortexes and tendexes for
this simulation.
\label{fig:SuperkickRemnant}
}
\end{figure}

To understand the mechanism of the recoil, consider the remnant black
hole just after merger.  The left panel of
Fig.~\ref{fig:SuperkickRemnant} shows the horizon tendicity and
the tendex lines emerging from the remnant
black hole at some particular time.  The
tendex structure is rotating counterclockwise around the hole's horizon.
The rotating tendex lines acquire accompanying vortex lines as
they extend into the wave zone in a pattern like that shown in
the right panel of
Fig.~\ref{fig:QNMKerr}, and there they become gravitational waves.
During merger, the horizon vortexes 
(right panel of Fig.~\ref{fig:SuperkickRemnant})
lock onto the same rotational angular velocity
as the horizon tendexes, and generate gravitational waves in
the same manner, with a pattern that looks like
the left panel of Fig.~\ref{fig:QNMKerr}. 

In the wave zone, the gravitational waves produced
by the rotating near-zone tendexes and those produced by the rotating
near-zone vortexes superpose coherently, and the resulting radiation pattern
depends on the angle between the 
horizon tendex labeled ``E'' and
the vortex labeled ``B'' 
in Fig.~\ref{fig:SuperkickRemnant}.  For the case shown, this
angle is 45 degrees, with $\mb E \times \mb B$ in the $-z$ direction
(into the page).
This is the same as the structure of a gravitational wave propagating in
the $-z$ direction (Fig.\ \ref{fig:PlaneWave}). As a result,
the gravitational waves produced by the vortexes and those
produced by the tendexes superpose constructively in the $-z$ direction
and destructively in the $+z$ direction, resulting in a maximum 
gravitational-wave 
momentum flow in the $-z$ direction and a maximum remnant
recoil in the $+z$ direction.

\subsubsection{Generic black-hole collisions}

\begin{figure}[b!]
\begin{center}
\includegraphics[width=0.95\columnwidth]{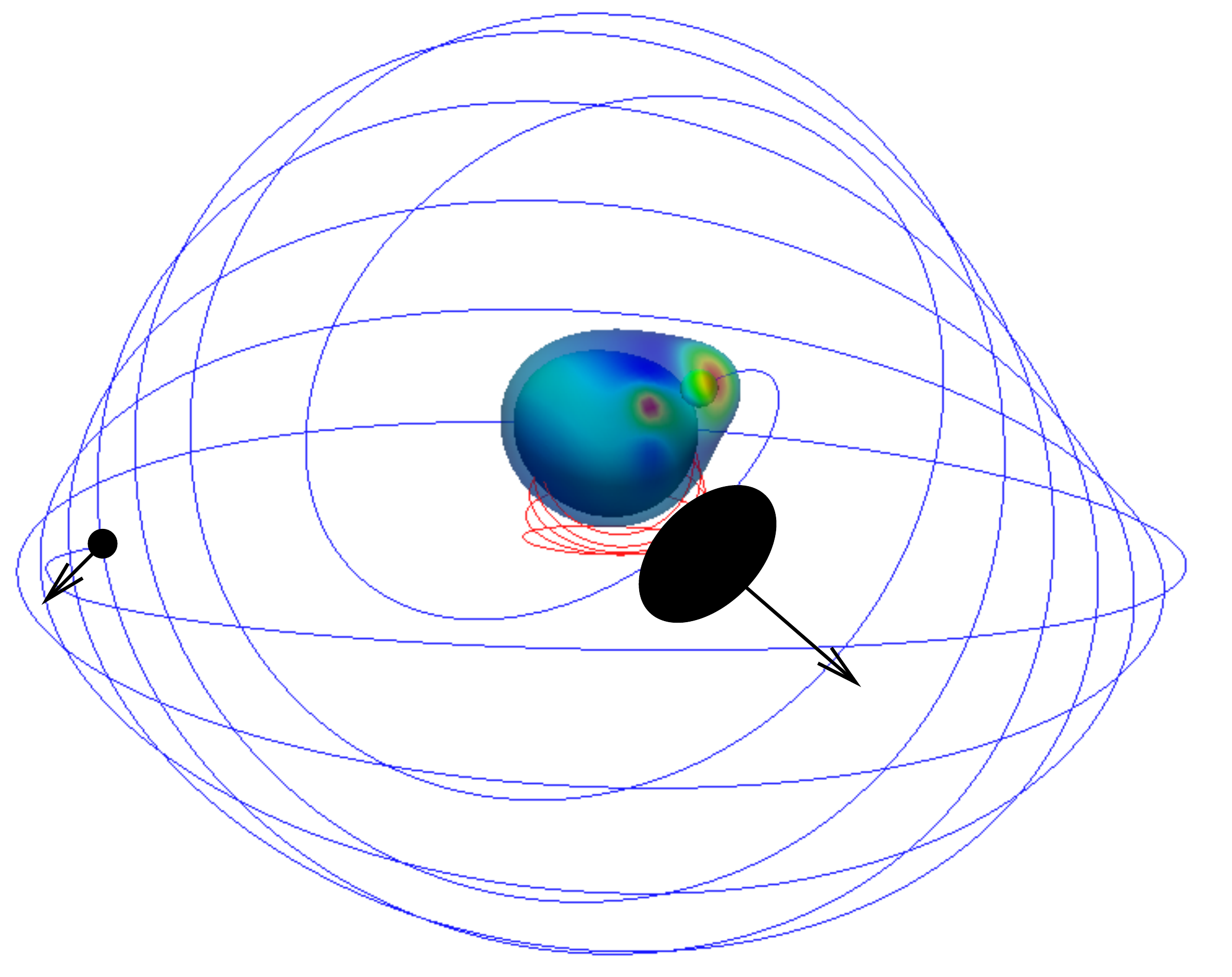}
\end{center}
\caption{(Color online.) The two thin 
curves are the trajectories of the centers
of two black holes in a 
generic orbit; from a simulation presented in~\cite{Mroue:2013PRL}.
The mass ratio of the two holes is 6:1, and the dimensionless spins
of the larger and smaller holes are  $S/M^2 = 0.91$ and $0.3$, respectively
(compared to the maximum possible spin $S/M^2=1$).
The initial black hole positions are shown in black, and the initial spins
are shown as arrows.
The spins are initially oriented in generic directions, so that the orbital
plane precesses.  Shown also are the apparent horizons of both holes,
and the common apparent horizon that encloses them, at the time the
common apparent horizon first forms.  The horizons are colored by the scalar
Ricci curvature, which is approximately $-2$ times the horizon tendicity.
\label{fig:GenericBBH}
See~\cite{movie:generic91} for a movie of this simulation.
}
\end{figure}

The geometrodynamic behaviors of more general black hole collisions are 
now being explored numerically.  For example, Fig.~\ref{fig:GenericBBH} shows
trajectories of two black holes in a fully generic orbit.  Vortexes
from the larger, rapidly-spinning black hole pull the orbit of the
smaller black hole into a complicated precession pattern.  The spin
directions of both black holes also precess as the holes orbit
each other.  Eventually a common apparent horizon\footnote{An apparent
horizon is a surface of zero outgoing null expansion, and lies inside or
on the event horizon.  Apparent horizons are local quantities that are
much easier to find in numerical simulations than are event horizons, because
the location of the event horizon depends on the full future evolution
of the spacetime.} forms around the two
individual apparent horizons, and the two holes merge into one.
The Ricci scalar (approximately equal to -2 times the horizon tendicity)
is shown on the two individual horizons and on the common horizon,
at the moment the common horizon forms.

\subsubsection{Tidal disruption of a neutron star by a spinning black hole}

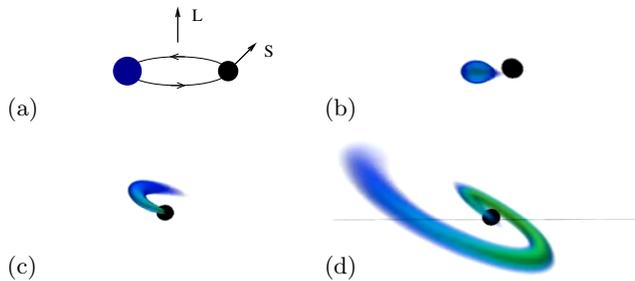
\begin{figure}[h!]
\begin{center}
\begin{picture}(0,110)(120,0)
\put(40,70){\includegraphics[width=0.25\columnwidth]{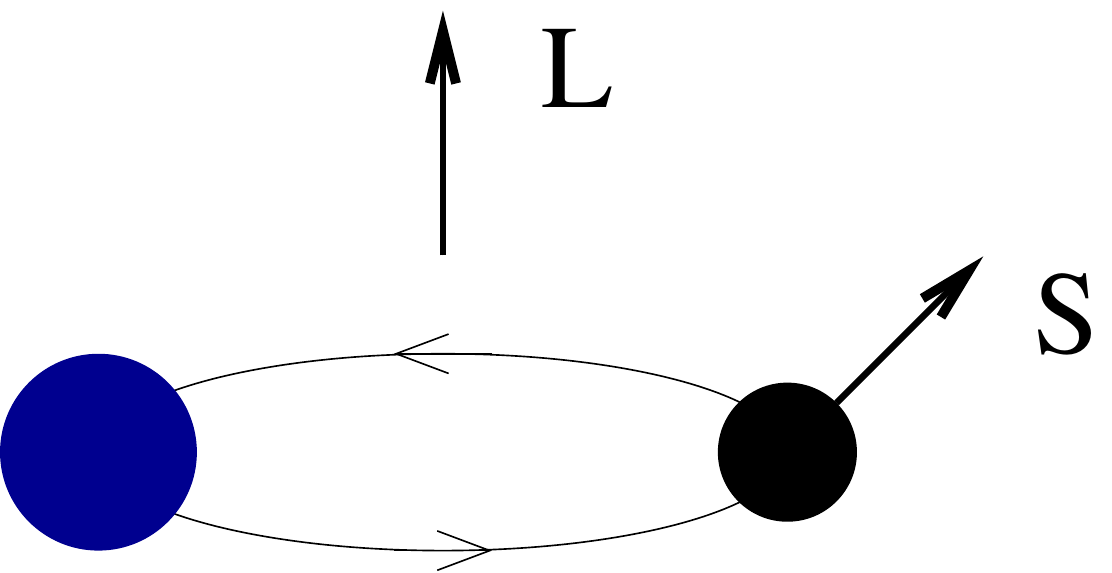}}
\put(0,60){(a)}
\put(150,60){
\includegraphics[width=0.25\columnwidth]{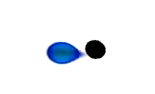}}
\put(120,60){(b)}
\put(0,-10){\includegraphics[width=0.47\columnwidth]{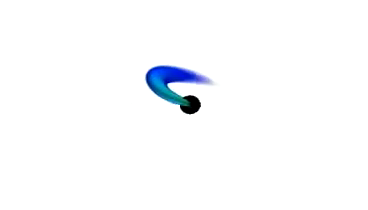}}
\put(0,0){(c)}
\put(120,-10){
\includegraphics[width=0.47\columnwidth]{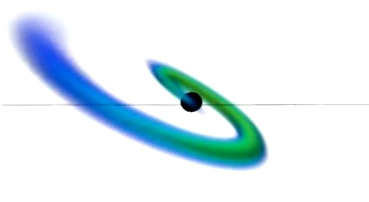}}
\put(120,0){(d)}
\end{picture}
\end{center}
\caption{(Color online.) Four successive snapshots of a
collision between a black hole (black) and a neutron star (blue or gray), viewed
edge-on in the initial orbital plane; from a simulation reported 
in~\cite{Foucart:2010eq}.
(a) The initial black-hole spin $S$ is inclined with respect to the
initial orbital angular momentum $L$.
(b) The black hole's tendexes begin to rip apart the neutron star.
(c) Some of the matter falls down the black hole, but some remains outside
the horizon. (d) The black hole's vortexes pull the remaining matter
into the black hole's equatorial plane, forming 
a disk and a tidal tail.  Figure adapted from~\cite{Foucart:2010eq}.
See~\cite{movie:bhns} for a movie of this simulation.
\label{fig:BhNs}
}
\end{figure}

Our final example illustrates the interaction of geometrodynamics and
matter.  Figure~\ref{fig:BhNs} shows a simulation of a neutron star
orbiting a black hole, a binary system important for gravitational-wave 
detectors and possibly for high-energy astrophysical phenomena such
as gamma-ray bursts.  Here the black hole has 3 times the
mass of the neutron star, and a dimensionless spin $S/M^2 = 0.5$ in a direction
inclined approximately 45 degrees to the orbital angular momentum.  When the orbit
has shrunk sufficiently, because of energy lost to gravitational radiation,
the black hole's tidal tendexes rip apart the neutron star, and
its frame-drag vortexes then pull the stellar debris out of its
original orbit and into the black hole's equatorial plane.  If the
black hole has a small enough spin or a large enough mass, the neutron
star is not disrupted, but instead is swallowed whole by the black 
hole~\cite{Foucart2012}.

\section{Gravitational-Wave Observations}
\label{sec:GW}

Geometrodynamics generically produces gravitational waves. 
We are entering an era in which these waves, generated by
sources in the distant universe, will be detected on Earth. 

A first generation of interferometric gravitational-wave detectors
has operated at sensitivities where it would require luck
to see waves.  
We were not lucky~\cite{Colaboration:2011np,Aasi:2012rja}.

\begin{figure*}
\begin{center}
\includegraphics[width=2.0\columnwidth]{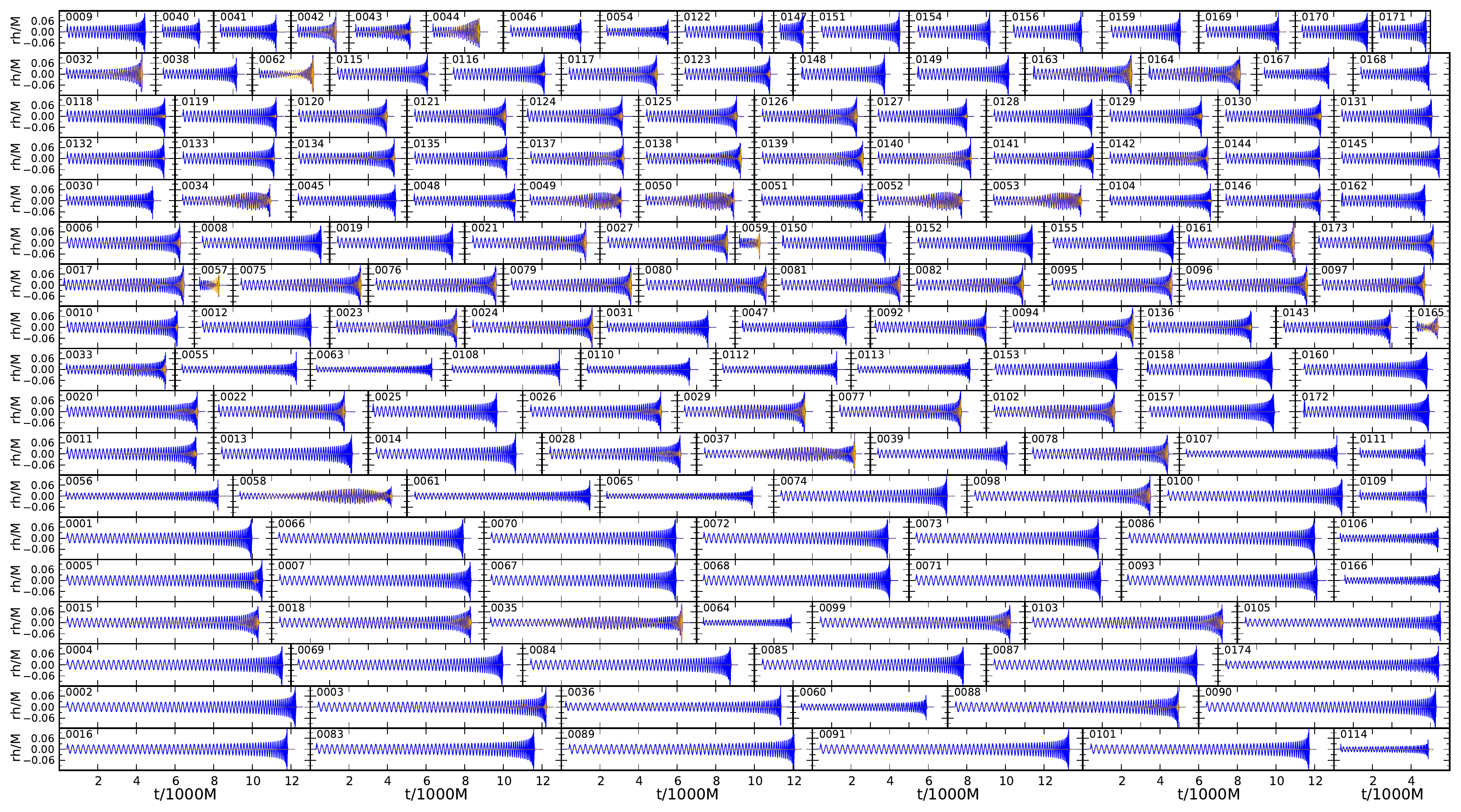}
\end{center}
\caption{(Color online.) A recent catalog of black-hole binary simulations
from~\cite{Mroue:2013PRL}. Shown are 174 waveforms, each with
two polarizations (shown in two colors or shades of gray)
in a sky direction parallel to
the initial orbital plane. The unit on the time axis corresponds to
0.1 s for a binary with a total mass of 20 solar masses.
}
\vskip-2pc
\label{fig:Catalog}
\end{figure*}

Second generation detectors, much more advanced and complex in their
design, are under construction. The first two of these 
(the advanced LIGO detectors in the U.S.) will begin
operating in 2015 and should reach their design sensitivity by
2019, and perhaps sooner~\cite{Aasi:2013wya}. They will be joined a bit 
later by the advanced VIRGO detector in Europe,
KAGRA in Japan, and an advanced LIGO detector in 
India~\cite{fairhurst:12}.
These advanced detectors will cover a 1000 times larger volume of
the universe than the initial detectors did, 
with estimated event rates for mergers of black-hole and neutron-star
binaries ranging from a few per year to a few
per week~\cite{DellerBailes2009,OShaughnessy2008,Abadie:2010cfa}.
Plans are being developed for further improvements, which should
increase the event rate by another factor ten or more, without major
changes in detector design.

The most interesting gravitational-wave sources for these detectors,
we think, are the dynamically evolving vortexes and tendexes attached
to merging black holes, and to a black hole tearing apart a neutron star
--- the geometrodynamic phenomena discussed above.  

The
numerical relativity community is building a catalog of binary simulations
and associated gravitational
waveforms to underpin the advanced detectors' searches for these waves.
Simulations of binary black holes with several hundred different sets 
of parameters (mass ratios  and initial vectorial spins) have been
completed \cite{Pekowsky:2013ska,Mroue:2013PRL,
Ajith:2012az,Ajith:2012az-PublicData,Ninja2_2014,Hinder:2013oqa};
and many more are underway or planned.  A sample of computed
waveforms from our SXS collaboration is shown in Fig.\  
\ref{fig:Catalog}.

Once waves are being detected, comparison of observed waveforms with
those from simulations will be crucial for understanding the waves' sources.
By such comparisons, can we deduce the sources' geometrodynamics and
test general relativity's predictions.

\section{Conclusions}
\label{sec:Conclusions}

Physicists have barely scratched the surface of geometrodynamics. As numerical
simulations continue to improve and are used to explore more
complicated and generic situations, we expect to learn more about the
geometrodynamics of critical behavior, singularities, dynamical
black holes, and other phenomena. We look forward to observations of 
gravitational waves
from strongly gravitating astrophysical sources, which will enable us
to test the geometrodynamical predictions of Einstein's equations for the
first time.

\begin{acknowledgments}
We are grateful to Thorne's dear friend, the late Leonid Petrovich Grishchuk, 
for urging us to
write this review article, and for many stimulating discussions.
We thank members of the SXS
collaboration and of the Caltech/Cornell Vortex/Tendex research
group for data used for figures in Sec~\ref{sec:BBH}, and for the
work that made much of that section possible.
We gratefully acknowledge support from the Sherman Fairchild
Foundation, the Brinson Foundation, and 
NSF grants PHY-106881 and AST-1333520.
\end{acknowledgments}

\bibliography{References/References}

\end{document}